\documentclass{article}
\usepackage[T1]{fontenc} 
\usepackage[utf8]{inputenc} 
\usepackage{ismir,amsmath,cite,url}
\usepackage{graphicx}
\usepackage{color}

\title{From West to East: Who can understand the music of the others better?}

\multauthor
{Charilaos Papaioannou$^{1,2}$ \hspace{1cm} Emmanouil Benetos$^2$ \hspace{1cm} Alexandros Potamianos$^1$}{
 $^1$ School of ECE, National Technical University of Athens, Greece\\
$^2$ Centre for Digital Music, Queen Mary University of London, UK\\
{\tt\small cpapaioan@mail.ntua.gr}
}

\def\authorname{C. Papaioannou, E. Benetos, and A. Potamianos}

\begin{document}

\maketitle
\begin{abstract}
Recent developments in MIR have led to several benchmark deep learning models whose embeddings can be used for a variety of downstream tasks. At the same time, the vast majority of these models have been trained on Western pop/rock music and related styles. This leads to research questions on whether these models can be used to learn representations for different music cultures and styles, or whether we can build similar music audio embedding models trained on data from different cultures or styles. To that end, we leverage transfer learning methods to derive insights about the similarities between the different music cultures to which the data belongs to. We use two Western music datasets, two traditional/folk datasets coming from eastern Mediterranean cultures, and two datasets belonging to Indian art music. Three deep audio embedding models are trained and transferred across domains, including two CNN-based and a Transformer-based architecture, to perform auto-tagging for each target domain dataset. Experimental results show that competitive performance is achieved in all domains via transfer learning, while the best source dataset varies for each music culture. The implementation and the trained models are both provided in a public repository.


\end{abstract}

\section{Introduction}\label{sec:intro}

As the time passes by, more and more pre-trained models are being made available in the MIR field. These models can be used in a variety of tasks by providing informative deep audio embeddings for music pieces. In correspondence with publicly available datasets, the vast majority of these models are trained on the so called ``Western''\footnote{we use the term ``Western'' to denote music styles which mostly originate from Western cultures, including pop, rock, and Western classical.} musical tradition \cite{gomez_computational_2013}. While studying world, folk, or traditional music, that fact arises two research questions; on the one hand what is the potential of these models when they are being used in the realm of a different culture and on the other hand how capable can a model be when trained on a specific music tradition on providing meaningful audio embeddings.

There are several experimental setups one can employ in order to derive answers to the above questions. By taking into account the importance of the auto-tagging task in the MIR field \cite{choi_deep_2018}, it becomes clear that transferring knowledge between domain-specific models to perform this task may lead us to valuable insights. Automatic content-based tagging aims to predict the tags of a music piece given its audio signal. The audio signal includes the acoustic characteristics and some of them are responsible for the occurrence of a tag in a piece, forming a multiple instance problem \cite{dietterich_solving_1997}.

A variety of models have been proposed to cope with the automatic tagging of music pieces. They can be divided, according to the input data they process, into the ones that utilize time-frequency representations and the others that accept the raw audio signal. In the first category, CNN-based models which are adopted by the computer vision field can be found, such as VGG-ish \cite{hershey_cnn_2017} as well as specifically developed architectures for music, like Musicnn \cite{pons_musicnn_2019}. A Transformer-based architecture was recently proposed in \cite{gong_ast_2021} called Audio Spectrogram Transformer (AST). Regarding the models that process audio, the TCNN \cite{pandey_tcnn_2019} and the Wave-U-Net \cite{stoller_wave-u-net_2018} architectures are being commonly used. For the purposes of our study, it is essential to use models of the same category with respect to the input they accept and, thus, we selected the ones that process time-frequency representations because of their popularity in the MIR field.

While using deep neural networks, transfer learning of a trained model can lead to a significant performance improvement on the target domain, compared to one that starts from a random state in the parameters space \cite{yang_transfer_2017}. Typically, the weights of the target domain model are initialized with the ones of a pre-trained model and then fine-tuning is applied. During this step, one has to determine which of the layers will be trainable and which ones will be kept frozen \cite{yosinski_how_2014}. In general, it is not clear which part of the network should be allowed to be trained in the target task and, thus, experimentation with different setups is necessary. Standard methods include the fine-tuning of the whole network, as suggested in \cite{girshick_rich_2014}, as well as only the last few layers or a part of the network, as in \cite{long_learning_2015}. We experiment with both setups to derive valuable insights on knowledge transfer across domains.

Even though under-represented in general, datasets from specific music cultures are evident in the MIR field and a set of the aforementioned methods have been used to perform several tasks. In \cite{sharma_classification_2021} a classification of Indian art music was conducted using deep learning models while automatic makam recognition in Turkish music was carried out in \cite{demirel_automatic_2018, ganguli2022critiquing}. With respect to Western music, there are several research works performing auto-tagging via deep learning models, as in \cite{choi_automatic_2016} and \cite{pons_end--end_2018}. 

In this paper, we incorporate a mosaic of different cultures by including six datasets from Western to Mediterranean and Indian music. Three music audio embedding models, two that mainly consist of convolutional layers and a Transformer-based architecture, are utilized on both single-domain and transfer learning experimental setups for music tagging. Results indicate that any model, despite the music culture that it is trained on, has the potential to adapt to another and achieve competitive results. When comparing the contributions of cross-domain knowledge transfers, we notice that they vary for each music culture and we suggest which one is the best candidate to outperform the single-domain approach. To the authors' knowledge, this is the first study which attempts to explore whether existing music audio embedding models can be used to transfer or learn representations for non-Western cultures. For reproducibility, we share the implementation in a public repository\footnote{https://github.com/pxaris/ccml}.

\section{Datasets}\label{sec:datasets}

The selection of the datasets is a prominent theme in the current study and it is constrained by the available corpora that reflect different music cultures. By basing our intuition on the location of each culture, we pursue to include three distinct geographic regions each one represented by two corpora. 

Even though spread in several continents, we consider the ``West'' as a single entity and utilize the MagnaTagATune \cite{magnatagatune_2009} and FMA-medium \cite{defferrard_fma_2017} datasets that mainly belong to this culture. The second region is the eastern Mediterranean represented by the traditions of Greece and Turkey in our study with Lyra \cite{papaioannou_dataset_2022} and Turkish-makam \cite{uyar_corpus_2014} datasets. The Indian subcontinent is also incorporated with Hindustani and Carnatic corpora \cite{srinivasamurthy_corpora_2014}, corresponding to the music traditions of the Northern and Southern areas of India respectively.

\begin{table*}[]
\begin{center}
\setlength\tabcolsep{1.2pt}
\begin{tabular}{|cc|cc|cc|cc|cc|cc|}
\hline
\multicolumn{2}{|c}{\textbf{MagnaTagATune}}                                                 & \multicolumn{2}{|c}{\textbf{FMA-medium}}                                                    & \multicolumn{2}{|c}{\textbf{Lyra}}                                                             & \multicolumn{2}{|c}{\textbf{Turkish-makam}}                                                              & \multicolumn{2}{|c}{\textbf{Hindustani}}                                                      & \multicolumn{2}{|c|}{\textbf{Carnatic}}                                                           \\ \hline
\multicolumn{1}{|l}{guitar}  & 18.76\% & \multicolumn{1}{l}{Rock}         & 28.41\% & \multicolumn{1}{l}{Voice}      & 76.21\% & \multicolumn{1}{l}{Voice}                & 63.33\% & \multicolumn{1}{l}{Voice}     & 83.90\% & \multicolumn{1}{l}{Voice}        & 82.35\% \\
\multicolumn{1}{|l}{classical}    & 16.52\% & \multicolumn{1}{l}{Electronic}   & 25.26\% & \multicolumn{1}{l}{Traditional}     & 76.05\% & \multicolumn{1}{l}{Kanun}                & 31.09\% & \multicolumn{1}{l}{Tabla}     & 53.03\% & \multicolumn{1}{l}{Violin}       & 78.45\% \\
\multicolumn{1}{|l}{slow}          & 13.71\% & \multicolumn{1}{l}{Punk}         & 13.28\% & \multicolumn{1}{l}{Violin}     & 57.34\% & \multicolumn{1}{l}{Tanbur}               & 27.93\% & \multicolumn{1}{l}{Khayal}          & 41.33\% & \multicolumn{1}{l}{Mridangam}    & 75.65\% \\
\multicolumn{1}{|l}{techno}       & 11.42\% & \multicolumn{1}{l}{Experimental} & 9.00\%  & \multicolumn{1}{l}{Percussion} & 53.71\% & \multicolumn{1}{l}{Ney}                  & 27.56\% & \multicolumn{1}{l}{Harmonium} & 39.25\% & \multicolumn{1}{l}{Kriti}              & 70.87\% \\
\multicolumn{1}{|l}{strings} & 10.55\% & \multicolumn{1}{l}{Hip-Hop}      & 8.80\%  & \multicolumn{1}{l}{Laouto}     & 51.69\% & \multicolumn{1}{l}{orchestra} & 26.38\% & \multicolumn{1}{l}{Teentaal}        & 35.35\% & \multicolumn{1}{l}{adi}                & 51.88\% \\
\multicolumn{1}{|l}{drums}   & 10.05\% & \multicolumn{1}{l}{Folk}         & 6.08\%  & \multicolumn{1}{l}{Guitar}     & 37.34\% & \multicolumn{1}{l}{Oud}                  & 24.36\% & \multicolumn{1}{l}{Tambura}   & 27.88\% & \multicolumn{1}{l}{Ghatam}       & 30.32\% \\
\multicolumn{1}{|l}{electronic}   & 9.74\%  & \multicolumn{1}{l}{Garage}       & 5.67\%  & \multicolumn{1}{l}{Klarino}    & 31.05\% & \multicolumn{1}{l}{kemence}    & 22.79\% & \multicolumn{1}{l}{Ektaal}          & 21.58\% & \multicolumn{1}{l}{Khanjira}     & 17.65\% \\
\multicolumn{1}{|l}{rock}         & 9.17\%  & \multicolumn{1}{l}{Instrumental} & 5.40\%  & \multicolumn{1}{l}{Nisiotiko}       & 26.85\% & \multicolumn{1}{l}{Cello}                & 17.83\% & \multicolumn{1}{l}{Pakhavaj}  & 7.88\%  & \multicolumn{1}{l}{rupaka}             & 11.98\% \\
\multicolumn{1}{|l}{fast}          & 8.92\%  & \multicolumn{1}{l}{Indie-Rock}   & 5.17\%  & \multicolumn{1}{l}{place-None}                                                       & 25.16\% & \multicolumn{1}{l}{Violin}               & 17.62\% & \multicolumn{1}{l}{Sarangi}   & 7.30\%  & \multicolumn{1}{l}{mishra chapu}       & 7.27\%  \\
\multicolumn{1}{|l}{piano}   & 7.95\%  & \multicolumn{1}{l}{Pop}          & 4.74\%  & \multicolumn{1}{l}{Bass}       & 24.76\% & \multicolumn{1}{l}{Hicaz}                     & 10.63\% & \multicolumn{1}{l}{Dhrupad}         & 7.05\%  & \multicolumn{1}{l}{Tana Varnam} & 5.21\%  \\ \hline
\end{tabular}
\end{center}
\caption{Relative frequencies of the top 10 most popular tags in each dataset.}
\label{table:top_tags}
\end{table*}

\subsection{MagnaTagATune}\label{subsec:magnatagatune}

MagnaTagATune \cite{magnatagatune_2009} is a publicly available dataset that is commonly used for the auto-tagging problem in the MIR field. It consists of more than 25,000 audio recordings, summing to 210 hours of audio content at total. Each audio recording is annotated with a subset of the unique 188 tags. Typically, only the top 50 most popular tags are used, which include annotations about genre, instruments and mood. In Table \ref{table:top_tags}, the most frequent tags for MagnaTagATune are presented along with the ones of the other datasets.

\subsection{FMA-medium}\label{subsec:fma}

The Free Music Archive \cite{defferrard_fma_2017} is an open and easily accessible dataset that is used for evaluating several tasks. It contains over 100,000 tracks which are arranged in a hierarchical taxonomy of 161 genres. In order to keep the durations of the datasets balanced whenever possible, and to include genres belonging to Western music styles, we use FMA-medium that consist of 25,000 tracks of 30 seconds each. That means that its total duration is 208 hours, almost equal to the one of MagnaTagATune. With regards to the metadata, we include the top-20 hierarchically related genres of the music pieces.

\subsection{Lyra}\label{subsec:lyra}

Lyra \cite{papaioannou_dataset_2022} is a dataset for Greek traditional and folk music that comprises 1570 pieces and metadata information with regards to instrumentation, geography and genre. Its total duration is 80 hours which makes it the only dataset with duration less than 200 hours in our study. We incorporate the top-30 tags retrieved from columns ``genre'', ``place'' and ``instruments'' to form our multi-label classification setup.

\subsection{Turkish-makam}\label{subsec:tmakam}

The Turkish makam corpus \cite{uyar_corpus_2014,senturk_computational_2016} includes thousands of audio recordings covering more than 2,000 works from hundreds of artists. It is part of CompMusic Corpora\footnote{https://compmusic.upf.edu/corpora} \cite{serra_creating_2014} which comprises data collections that have been created with the aim of studying particular music traditions. Using Dunya \cite{porter2013dunya} and the related software tool\footnote{https://github.com/MTG/pycompmusic}, we were able to get access to 5297 audio recordings, summing in 359 hours, along with their metadata. In order to keep the dataset sizes similar, we set a maximum audio duration equal to 150 seconds which reduced the total length to 215 hours. For the tags, the top-30 most popular with regards to ``makam'', ``usul'' and ``instruments'' information have been included.

\subsection{Hindustani}\label{subsec:shindustani}

The Hindustani corpus \cite{srinivasamurthy_corpora_2014} is also part of CompMusic Corpora. It includes 1204 audio recordings, with a total duration of 343 hours, covering a plethora of artists and metadata categories. By setting the maximum audio duration to 780 seconds, the size of the dataset has been decreased to 206 hours for the needs of our study. Furthermore, information about ``raga'', ``tala'', ``instruments'' and ``form'' has been used to form the labels of each piece. The top-20 most frequent tags have been incorporated to our study as the target of the classification models. 

\subsection{Carnatic}\label{subsec:scarnatic}

The Carnatic corpus \cite{srinivasamurthy_corpora_2014} comprises 2612 audio recordings, summing in more than 500 hours of content. As with the previous datasets, by setting a maximum duration cut equal to 330 seconds, the total duration has been decreased to 218 hours. Identical to Hindustani, the top-20 most popular annotations regarding ``raga'', ``tala'', ``instruments'' and ``form'' have been included for the metadata.

\section{Method}\label{sec:method}

In this section, the models which are used for the purposes of this study are first presented. We, then, describe how transfer learning is utilized to infer similarities between the music cultures by employing knowledge from the domain adaptation field. 

\subsection{Models}

\subsubsection{VGG-ish}

All of our models use the mel-spectrogram as their input, a commonly used feature for MIR tasks such as automatic tagging \cite{dieleman_multiscale_nodate}. This selection enables the utilization of CNN-based architectures which have been successfully used in computer vision tasks. The Visual Geometry Group (VGG) network \cite{simonyan_very_2015} and its variants consist of a stack of convolutional layers followed by fully connected layers.

We use a VGG-ish architecture, similar to the one implemented by the authors in \cite{won_evaluation_2020}, that is a 7-layer CNN, with $3\times3$ convolution filters and $2\times2$ max-pooling, followed by two fully-connected layers. It accepts mel-spectrograms that correspond to short chunks of audio as its input, with duration equal to 3.69 seconds.

\subsubsection{Musicnn}

Musicnn \cite{pons_end--end_2018} is a music inspired model that uses convolutional layers at its core. Its first convolutional layer consists of vertical and horizontal filters in order to capture timbral and temporal features respectively. These features are, then, concatenated and fed to 1D convolutional layers followed by a pair of dense layers that summarize them and predict the relevant tags. Similar to VGG-ish, it uses mel spectrograms from short audio chunks at its input with duration 3 seconds.

\subsubsection{Audio Spectrogram Transformer}

As its name indicates, Audio Spectrogram Transformer (AST) is a purely attention-based model for audio classification \cite{gong_ast_2021}. Based on the Transformer architecture \cite{vaswani_attention_2017}, AST splits the input mel-spectrogram to $16\times16$ patches in both time and frequency dimensions that are, in turn, flattened to 1D embeddings of size 768 using a linear projection layer. A trainable positional embedding is also added to each patch embedding so that the model will capture the spatial structure of the input 2D spectrogram. The resulting sequence is fed to the Transformer, where only the encoder is utilized since AST is designed for classification tasks. The output of the encoder is followed by a linear layer that predicts the labels. As the authors that introduced the architecture suggest, we set a specific cut to the input length of the AST model that is equal to 8 seconds in all our experiments.

\begin{table*}[t!]
    \begin{center}
        \begin{tabular}{ccccccc}
            \hline
            Model                                                      & \multicolumn{2}{c}{\textbf{VGG-ish}}          & \multicolumn{2}{c}{\textbf{Musicnn}}          & \multicolumn{2}{c}{\textbf{AST}}              \\ \hline
            \begin{tabular}[c]{@{}c@{}}Metric /\\ Dataset\end{tabular} & \multicolumn{1}{c}{ROC-AUC} & PR-AUC & \multicolumn{1}{c}{ROC-AUC} & PR-AUC & \multicolumn{1}{c}{ROC-AUC} & PR-AUC \\ \hline
            \textbf{MagnaTagATune}                                              & \multicolumn{1}{c}{0.9123}       & 0.4582      & \multicolumn{1}{c}{0.9019}       & 0.4333      & \multicolumn{1}{c}{\textbf{0.9172}}       & \textbf{0.4654}      \\
            \textbf{FMA-medium}                                                        & \multicolumn{1}{c}{\textbf{0.8889}}       & 0.4949      & \multicolumn{1}{c}{0.8766}       & 0.4473      & \multicolumn{1}{c}{0.8886}       & \textbf{0.5024}      \\
            \textbf{Lyra}                                                       & \multicolumn{1}{c}{0.8097}       & 0.4806      & \multicolumn{1}{c}{0.7391}       & 0.4042      & \multicolumn{1}{c}{\textbf{0.8476}}       & \textbf{0.5333}      \\
            \textbf{Turkish-makam}                                              & \multicolumn{1}{c}{\textbf{0.8696}}       & 0.5639      & \multicolumn{1}{c}{0.8505}       & 0.5299      & \multicolumn{1}{c}{0.8643}       & \textbf{0.5669}      \\
            \textbf{Hindustani}                                                 & \multicolumn{1}{c}{\textbf{0.8477}}       & \textbf{0.6082}      & \multicolumn{1}{c}{0.8471}       & 0.6016      & \multicolumn{1}{c}{0.8307}       & 0.5786      \\
            \textbf{Carnatic}                                                   & \multicolumn{1}{c}{0.7392}       & 0.4278      & \multicolumn{1}{c}{0.7496}       & 0.4182      & \multicolumn{1}{c}{\textbf{0.7706}}       & \textbf{0.4394}      \\ \hline
        \end{tabular}
    \end{center}
    \caption{ROC-AUC and PR-AUC scores of the models on single domain auto-tagging tasks.}
\label{table:single_domain}
\end{table*}
 
\subsection{Transfer Learning}

The purpose of transfer learning is to improve the performance of the models on target domains by transferring knowledge from different but related source domains \cite{pan_survey_2010}. In the field of MIR, both transferring feature representations to the target domain from a pre-trained model on a source task \cite{van_den_oord_transfer_2014} as well as learning shared latent representations across domains \cite{hamel_transfer_2013} have been proposed in the past. Yet, these methods have not been applied to non-Western music datasets neither by adapting an existing model to them nor by studying to what end these cultures can be valuable source domains for widely developed models, two aspects which are both studied in this work.

According to the categorization conducted by the authors in \cite{zhuang_comprehensive_2020}, these methods belong to \textit{parameter sharing} category of the model-based transfer learning techniques. In the deep learning realm, it is, thus, common to use a trained network for a source task, share its parameters and in turn fine-tune some or all  layers to produce a target network. While following this method, one expects it to lead to better results when the participating domains are similar to each other. Indeed, by studying the prior work on domain adaptation, one will find that the main strategy consists of minimizing the difference between the source and target feature distributions, when transferring representations from a labeled dataset to a target domain where labeled data is sparse or non-existent \cite{tzeng_simultaneous_2015,saenko_adapting_2010}.

By adapting the above rationale to our study, where the participating domains are all rich in labeled data, we expect that when applying transfer learning by parameter sharing, the more the similarity between the participating domains the better the performance of the target domain on its supervised learning task. 

In order to study to what end this hypothesis stands in computational musicology with deep neural networks, we utilize the previously presented models which are widely used in the MIR field and consist of different cores, namely convolutional layers (VGG-ish and Musicnn) and a Transformer module (AST). Having the models trained on each single dataset, we apply all the cross-domain knowledge transfers for each architecture by fine-tuning only the output layer as well as the whole network. We then aggregate the results across the models seeking to derive insights with regards to the similarities between the domains as well as specifying which source is the best candidate for each target dataset.

\section{Experiments}\label{sec:experiments}

As already mentioned, we use mel spectrograms as the input of all our models. In order to convert the audio recordings of the datasets to this representation, we use Librosa \cite{mcfee_librosa_2015} to re-sample them to 16 kHz sample rate. Then, 512-point FFT with a 50\% overlap is applied, the maximum frequency is set to 8 kHz and number of Mel bands to 128. Our intention, in this study, is not the optimization of the performance of the single-domain tasks but rather studying the knowledge transfer across the domains. So, we keep our training setup as close as possible to the literature, at each single domain task, in order to have a sanity check for the implementation. 

For VGG-ish and Musicnn models, we use a mixture of scheduled Adam \cite{kingma_adam_2017} and stochastic gradient descent (SGD) for the optimization method, identical to what the authors at \cite{won_evaluation_2020} have used. The batch size is set to 16 and the learning rate to $1e-4$ for both models while the maximum number of epochs are 200 for VGG-ish and 50 for Musicnn. With regards to the AST model, we follow the setup proposed in \cite{gong_ast_2021}, namely batch size 12, Adam optimizer, learning rate scheduling that begins from $1e-5$ and is decreased by a factor of 0.85 every epoch after the 5th one as well as pre-trained on Imagenet Transformer weights.

All our models accept a fixed size audio chunk at their input but need to predict song-level tags. During the evaluation phase, we aggregate the tag scores across all chunks by averaging them to acquire the label scores for the whole audio. We use the area under receiver operating characteristic curve (ROC-AUC), a widely used evaluation metric on multi-label classification problems and the area under precision-recall curve (PR-AUC), a suitable metric for unbalanced datasets \cite{davis_relationship_2006}.

During transfer learning, we initialize all parameters of the target model, except for the output layer, from each source dataset and (i) allow only the output layer to be trained and (ii) train the whole network. In both settings, we use the same hyper-parameters and evaluation procedure with the single-domain setups across all datasets for each model architecture.

\section{Results}\label{sec:results}

\begin{table*}[h!]
    \begin{center}
    \setlength\tabcolsep{3.2pt}
        \begin{tabular}{ccccccccccccc}
        \hline        
        Target domain                                                                & \multicolumn{2}{c}{\textbf{MagnaTagATune}} & \multicolumn{2}{c}{\textbf{FMA-medium}} & \multicolumn{2}{c}{\textbf{Lyra}} & \multicolumn{2}{c}{\textbf{Turkish-makam}} & \multicolumn{2}{c}{\textbf{Hindustani}} & \multicolumn{2}{c}{\textbf{Carnatic}} \\ \hline
        \begin{tabular}[c]{@{}c@{}}trainable layer(s) /\\ Source domain\end{tabular} & \multicolumn{1}{c}{output}      & all      & \multicolumn{1}{c}{output}     & all    & \multicolumn{1}{c}{output}  & all & \multicolumn{1}{c}{output}      & all      & \multicolumn{1}{c}{output}     & all    & \multicolumn{1}{c}{output}    & all   \\ \hline
        \multicolumn{13}{c}{\textbf{VGG-ish}}\\ \hline
        \textbf{MagnaTagATune}                                                       & \multicolumn{1}{c}{-}           & 91.23        & \multicolumn{1}{c}{88.11}          & \textbf{92.39}      & \multicolumn{1}{c}{74.69}       & \textbf{85.40}   & \multicolumn{1}{c}{76.79}           & 86.84        & \multicolumn{1}{c}{76.09}          & 85.04     & \multicolumn{1}{c}{67.19}        & \textbf{74.71}    \\
        \textbf{FMA-medium}                                                          & \multicolumn{1}{c}{85.82}           & \textbf{91.29}        & \multicolumn{1}{c}{-}          & 88.89      & \multicolumn{1}{c}{68.56}       & 84.04   & \multicolumn{1}{c}{75.40}           & \textbf{87.78}        & \multicolumn{1}{c}{75.77}          & 84.39     & \multicolumn{1}{c}{67.03}        & 74.56    \\
        \textbf{Lyra}                                                                & \multicolumn{1}{c}{84.34}           & 90.93        & \multicolumn{1}{c}{82.84}          & 92.10      & \multicolumn{1}{c}{-}       & 80.97   & \multicolumn{1}{c}{76.98}           & 87.21        & \multicolumn{1}{c}{77.41}          & 84.24     & \multicolumn{1}{c}{67.30}        & 73.52    \\
        \textbf{Turkish-makam}                                                       & \multicolumn{1}{c}{85.19}           & 90.90        & \multicolumn{1}{c}{84.41}          & 91.74      & \multicolumn{1}{c}{70.93}       & 82.38   & \multicolumn{1}{c}{-}           & 86.96        & \multicolumn{1}{c}{77.54}          & \textbf{85.32}     & \multicolumn{1}{c}{67.16}        & 73.50    \\
        \textbf{Hindustani}                                                          & \multicolumn{1}{c}{84.24}           & 91.02        & \multicolumn{1}{c}{83.83}          & 91.91      & \multicolumn{1}{c}{66.27}       & 79.71   & \multicolumn{1}{c}{77.25}           & 87.63        & \multicolumn{1}{c}{-}          & 84.77     & \multicolumn{1}{c}{66.72}        & 74.63    \\
        \textbf{Carnatic}                                                            & \multicolumn{1}{c}{84.18}           & 91.00        & \multicolumn{1}{c}{82.62}          & 91.73      & \multicolumn{1}{c}{61.59}       & 76.72   & \multicolumn{1}{c}{77.07}           & 87.40        & \multicolumn{1}{c}{78.19}          & 84.81     & \multicolumn{1}{c}{-}        & 73.92    \\ \hline
        \multicolumn{13}{c}{\textbf{Musicnn}}\\ \hline
        \textbf{MagnaTagATune}                                                       & \multicolumn{1}{c}{-}           & 90.19        & \multicolumn{1}{c}{87.34}          & \textbf{91.03}      & \multicolumn{1}{c}{71.79}       & 78.74   & \multicolumn{1}{c}{74.72}           & \textbf{85.96}        & \multicolumn{1}{c}{75.87}          & 84.18     & \multicolumn{1}{c}{66.12}        & 75.57    \\
        \textbf{FMA-medium}                                                          & \multicolumn{1}{c}{85.52}           & \textbf{90.35}        & \multicolumn{1}{c}{-}          & 87.66      & \multicolumn{1}{c}{65.94}       & 77.59   & \multicolumn{1}{c}{75.51}           & 85.13        & \multicolumn{1}{c}{73.16}          & \textbf{85.49}     & \multicolumn{1}{c}{66.38}        & 75.77    \\
        \textbf{Lyra}                                                                & \multicolumn{1}{c}{81.38}           & 90.03        & \multicolumn{1}{c}{82.23}          & 90.80      & \multicolumn{1}{c}{-}       & 73.91   & \multicolumn{1}{c}{74.11}           & 85.20        & \multicolumn{1}{c}{78.10}          & 83.29     & \multicolumn{1}{c}{65.09}        & 75.51    \\
        \textbf{Turkish-makam}                                                       & \multicolumn{1}{c}{84.35}           & 90.11        & \multicolumn{1}{c}{83.79}          & 90.81      & \multicolumn{1}{c}{61.87}       & \textbf{79.83}   & \multicolumn{1}{c}{-}           & 85.05        & \multicolumn{1}{c}{75.67}          & 83.75     & \multicolumn{1}{c}{67.49}        & 74.09    \\
        \textbf{Hindustani}                                                          & \multicolumn{1}{c}{82.38}           & 89.86        & \multicolumn{1}{c}{83.42}          & 90.85      & \multicolumn{1}{c}{64.48}       & 78.95   & \multicolumn{1}{c}{74.60}           & 85.58        & \multicolumn{1}{c}{-}          & 84.71     & \multicolumn{1}{c}{65.25}        & \textbf{76.95}    \\
        \textbf{Carnatic}                                                            & \multicolumn{1}{c}{83.02}           & 90.05        & \multicolumn{1}{c}{82.78}          & 90.74      & \multicolumn{1}{c}{61.83}       & 77.92   & \multicolumn{1}{c}{75.09}           & 85.43        & \multicolumn{1}{c}{75.34}          & 84.19     & \multicolumn{1}{c}{-}        & 74.96    \\ \hline
        \multicolumn{13}{c}{\textbf{AST}}\\ \hline
        \textbf{MagnaTagATune}                                                       & \multicolumn{1}{c}{-}           & \textbf{91.72}        & \multicolumn{1}{c}{89.25}          & 91.99      & \multicolumn{1}{c}{75.68}       & 83.77   & \multicolumn{1}{c}{76.28}           & 87.20        & \multicolumn{1}{c}{74.67}          & \textbf{86.57}     & \multicolumn{1}{c}{66.03}        & 75.43    \\
        \textbf{FMA-medium}                                                          & \multicolumn{1}{c}{88.63}           & 91.62        & \multicolumn{1}{c}{-}          & 88.86      & \multicolumn{1}{c}{65.72}       & 82.17   & \multicolumn{1}{c}{76.37}           & \textbf{87.43}        & \multicolumn{1}{c}{74.51}          & 85.76     & \multicolumn{1}{c}{67.33}        & 75.98    \\
        \textbf{Lyra}                                                                & \multicolumn{1}{c}{87.49}           & 91.44        & \multicolumn{1}{c}{87.44}          & \textbf{92.43}      & \multicolumn{1}{c}{-}       & \textbf{84.76}   & \multicolumn{1}{c}{77.08}           & 86.80        & \multicolumn{1}{c}{72.24}          & 83.73     & \multicolumn{1}{c}{68.47}        & 76.59    \\
        \textbf{Turkish-makam}                                                       & \multicolumn{1}{c}{87.33}           & 91.40        & \multicolumn{1}{c}{86.31}          & 91.95      & \multicolumn{1}{c}{72.70}       & 77.95   & \multicolumn{1}{c}{-}           & 86.43        & \multicolumn{1}{c}{70.13}          & 83.56     & \multicolumn{1}{c}{67.10}        & 75.23    \\
        \textbf{Hindustani}                                                          & \multicolumn{1}{c}{87.40}           & 91.35        & \multicolumn{1}{c}{87.11}          & 92.26      & \multicolumn{1}{c}{71.74}       & 84.60   & \multicolumn{1}{c}{75.70}           & 86.90        & \multicolumn{1}{c}{-}          & 83.07     & \multicolumn{1}{c}{67.75}        & 75.85    \\
        \textbf{Carnatic}                                                            & \multicolumn{1}{c}{87.42}           & 91.45        & \multicolumn{1}{c}{86.83}          & 91.75      & \multicolumn{1}{c}{63.33}       & 81.44   & \multicolumn{1}{c}{76.87}           & 87.14        & \multicolumn{1}{c}{74.11}          & 82.91     & \multicolumn{1}{c}{-}        & \textbf{77.06}    \\ \hline        
        \end{tabular}
        \caption{ROC-AUC scores (\%) when applying transfer learning using the models VGG-ish, Musicnn and Audio Spectrogram Transformer. Rows are the source domains and columns the target domains. After initializing the network with the parameters of the trained (at the source dataset) model, fine-tuning on the output layer as well as on the whole network is applied. The diagonal values (under the ``all'' columns) correspond to the respective single-domain models (no transfer learning) where the experimentation with only the output layer trainable has no meaning.}
        \label{table:transfer_learning}
    \end{center}
\end{table*}

The performance of the three models on all single-domain tasks can be seen in Table \ref{table:single_domain}. The performance of the Musicnn and VGG-ish models on MagnaTagATune is similar to the reported metrics in \cite{won_evaluation_2020}, which indicates the validity of our implementation. In general, the AST model shows the best performance followed by VGG-ish and then Musicnn. This result should not be taken into account solidly, because no hyper-parameter tuning has been taken place for each domain and in order to keep the duration of the training to less than 24 hours for each task, the number of epochs for Musicnn was significantly less than VGG-ish. On the other hand, one should consider that the AST \cite{gong_ast_2021} and VGG-ish \cite{won_evaluation_2020} models may, indeed,  perform better for limited time resources.

In Table \ref{table:transfer_learning}, one can see the ROC-AUC scores in all single-domain and cross-domain setups. The rows are the source datasets while the columns are the target datasets. A sub-table is constructed for each model architecture and for a transfer from domain \textit{A} to \textit{B}, the result of the fine-tuning of only the output layer (`output') as well as all the layers (`all') are reported. The single-domain setup is when source and target is the same dataset and, thus, only training of the whole network has meaning. The table is better parsed column-wise, e.g., by inspecting the results of VGG-ish model on MagnaTagATune when transferring knowledge from the other domains at the upper-left pair of columns in the table.

\begin{figure}[t]
 \centerline{
 \includegraphics[width=\columnwidth]{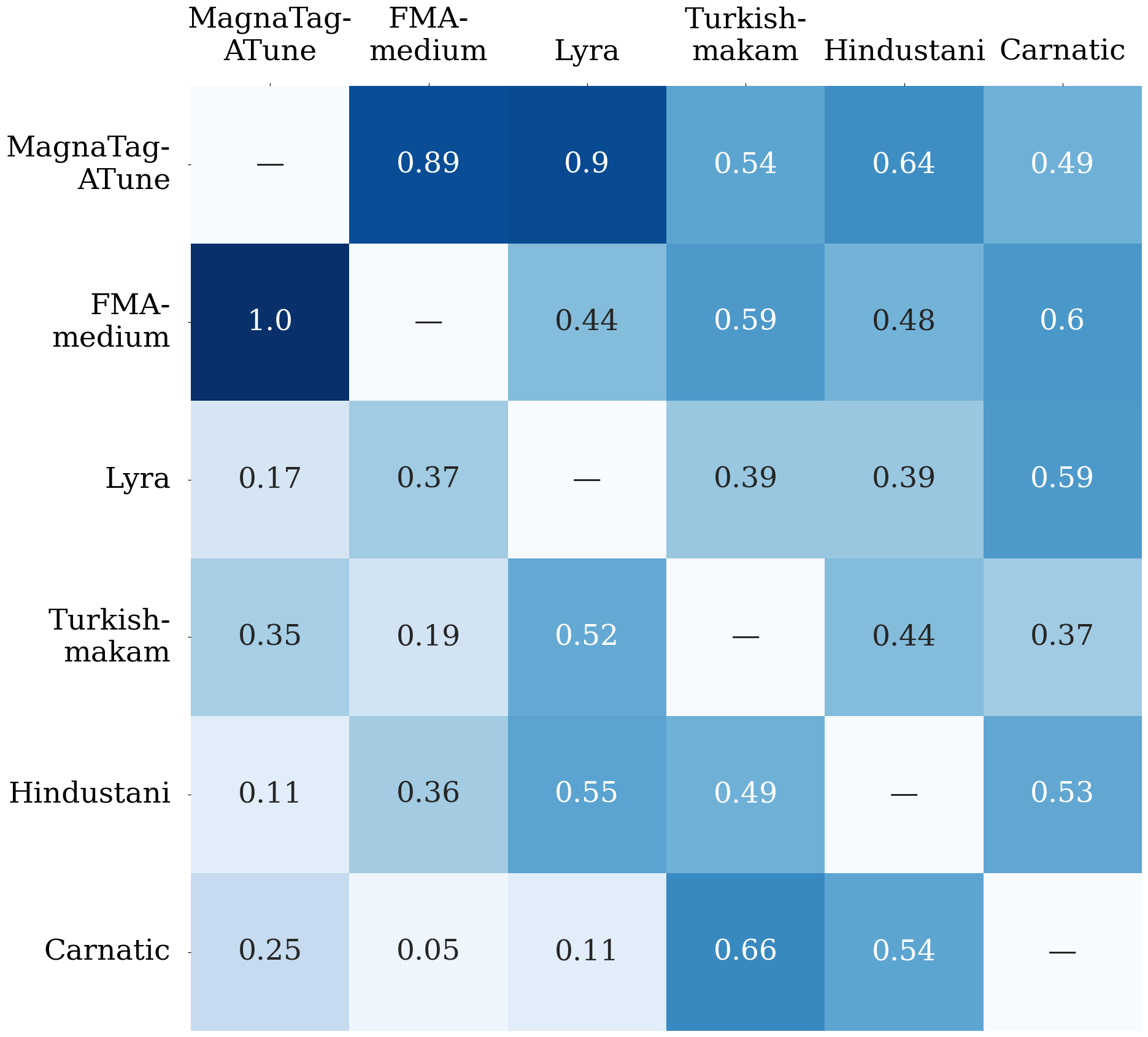}}
 \caption{Cross-cultural music transfer learning results. Rows correspond to the source datasets and columns to the target datasets. The value of each cell (knowledge transfer) is normalized and averaged across all models and fine-tuning methods.}
 \label{fig:cm_cultures}
\end{figure}

In order to aggregate all the cross-domain knowledge transfers, we follow the subsequent procedure:
for each target task that consists of a specific model, target dataset and fine-tuning method, min-max normalization is applied to the $N-1$ transfer learning results, where $N$ is the number of all datasets. The previous step leads to the construction of $M \times F$ matrices, $M$ the number of the models and $F$ the number of fine-tuning methods, where rows are the source domains, columns the target domains and diagonal elements are empty. Each cell has a value in the range $[0, 1]$, as a result of the normalization step, while the value $1$ corresponds to the knowledge transfer that led to the best performance in the target domain. 
By calculating the element-wise mean of the produced $M \times F$ matrices, we reach to the result that can be seen in Figure \ref{fig:cm_cultures}.

\begin{figure*}[t]
 \centerline{
 \includegraphics[width=0.83\paperwidth]{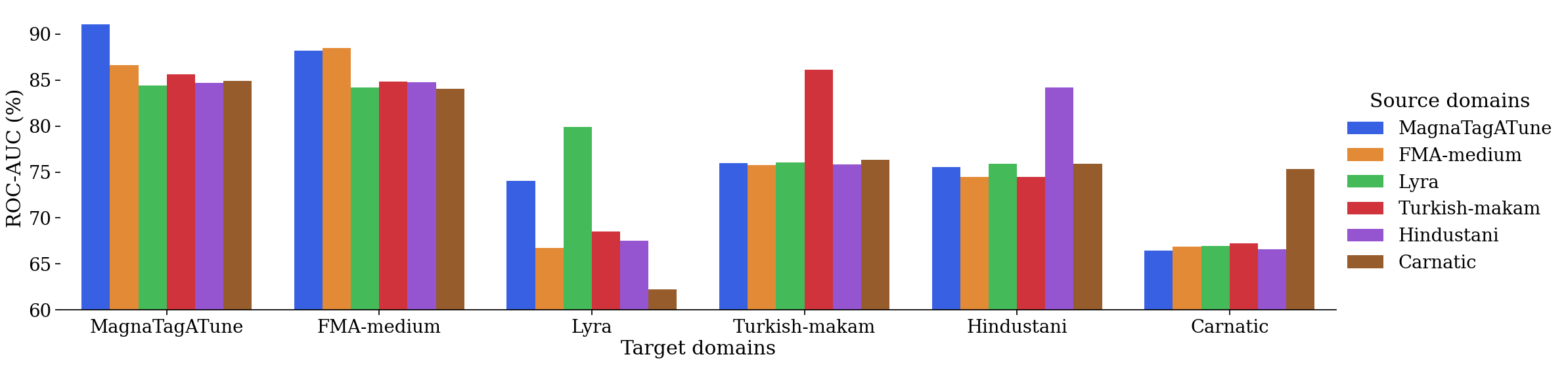}}
 \caption{Average, over the three models, ROC-AUC scores of all cross-domain transfers when fine-tuning of the output layer is applied. The highest bar at each group corresponds to the respective single-domain model.}
 \label{fig:barplot}
\end{figure*}

\section{Discussion}\label{sec:discussion}

The results indicate that knowledge transfer both from Western to non-Western cultures and the opposite can be beneficial when deep learning models are used to perform automatic music tagging. Indeed, by inspecting Table \ref{table:transfer_learning}, the general take-home message one should acquire is that regardless of the model architecture, all datasets have the potential to contribute as a source to a target domain by providing their deep audio embeddings. To investigate how valuable knowledge transfers from widely used datasets to non-Western music cultures can be, we focus on the last four datasets, i.e., the last eight columns of the table, and parse the two first rows, corresponding to MagnaTagATune and FMA datasets, at each model architecture. For instance, we notice that for Lyra, when Musicnn is used and fine-tuning only of the output layer is applied, the model coming from MagnaTagATune has the greater ROC-AUC score, namely 71.79\%. Additionally, the AST model trained on the FMA-medium dataset, outperforms the others when totally fine-tuned to the Turkish-makam dataset, scoring 87.43\%.

In order to study the inverse transfer direction, we center our interest to the first four columns of the entire table. Even though MagnaTagATune and FMA are almost always the best source for each other, the deep audio embeddings provided by the other datasets achieve competitive performance. For example, when MagnaTagATune is the target domain and fine-tuning is restricted to the output layer of the network, we observe that transferring from Turkish-makam leads to a performance that is comparable to the best source (FMA-medium) for all models.

By considering all cross-domain knowledge transfers, one can specify the best candidate to provide a trained model, with a specific architecture, for each target dataset. We, thus, notice that the model that is transferred from Hindustani outperforms the others at the Carnatic dataset, when fine-tuning on the whole Musicnn architecture is applied. A holistic picture of the cross-cultural music transfer learning is depicted in Figures \ref{fig:barplot} and \ref{fig:cm_cultures}. 

In Fig.~\ref{fig:barplot} the scores of all cross-domain transfers when fine-tuning the output layer, can be seen, averaged across the three models. The uniformity of the performances of different sources at each target dataset can be examined. We, thus, recognize that the most unbalanced performances are spotted on the Lyra target domain, a result that is probably related to the smaller size of this dataset compared to the others. By exploring Fig.~\ref{fig:cm_cultures} in a column-wise fashion, we observe that for MagnaTagATune as the target domain, FMA-medium is the best source with a value equal to $1$. This means that in all transfer learning setups, this source performed better than the others in this domain.

Both figures show that MagnaTagATune and FMA-medium perform consistently well across the domains, something that possibly indicates their appropriateness for the auto-tagging task. However, as we move to the Eastern cultures, we notice that their contribution is somehow decreased and other domains tend to contribute similarly or even more in those targets. The values at  Fig.~\ref{fig:cm_cultures} should not be considered solidly as similarity metrics between the domains because other factors may also affect the results we notice. It is, although, a first step towards studying different music cultures using deep learning methods.

\section{Conclusions}\label{sec:conclusions}

In this paper, the transferrability of music cultures by utilizing deep audio embedding models is studied. To that end, six datasets and three models were employed while experimentation with two fine-tuning methods took place. The automatic tagging of music pieces served as the supervised learning task where all cross-domain knowledge transfers were applied and evaluated.

The results show that state-of-the-art models can benefit from knowledge transfer not only from Western to non-Western cultures but also the opposite too. By aggregating the scores across all models and fine-tuning methods, the suitability of each source domain for a target task was calculated and, thus, which domain can be the best candidate to transfer knowledge from for each dataset was proposed. Based on the literature, we suggest that this result can be interpreted to a degree as a similarity metric between the music cultures.

We identify that the current study has limitations. In the future, the semantic similarities between the labels of the involved domains will be examined. More datasets and models, like those that process raw audio signals, will be considered as well as semi-supervised and unsupervised learning techniques. Other tasks may be employed such as mode estimation, assuming that key in Western cultures functions in a similar way with makam or raga in other cultures. All datasets can also be utilized to learn music embeddings in order to unveil cross-cultural links between acoustic features and tags.

\section{Acknowledgements}\label{sec:acknowledgements}

The authors would like to thank Sertan {\c{S}}ent{\"u}rk, Alastair Porter and the Universitat Pompeu Fabra for their willingness to provide us with the data without which this study would not have been possible. We would like to also thank Charalampos Saitis and the reviewers for their valuable and constructive comments that helped us improve our work.

\bibliography{ISMIRtemplate}

%
%
%
%
%

\end{document}